**Title:** Dosimetric Quantification of a Commercial Dual-Tube kV X-Ray System for Preclinical FLASH Research

**Authors:**

Luka Matej Devenica[1†], Lixiang Guo[1†], Mohammad Rezaee[2] and Ken Kang-Hsin Wang[1*]

[†]These authors contributed equally to this work.
[1]Biomedical Imaging and Radiation Technology Laboratory (BIRTLab), Department of Radiation Oncology, University of Texas Southwestern Medical Center, Dallas, Texas, USA
[2]Department of Radiation Oncology and Molecular Radiation Sciences, Johns Hopkins University, Baltimore, MD, USA

***Correspondence:**

Ken Kang-Hsin Wang, PhD

Department of Radiation Oncology, University of Texas Southwestern Medical Center, 5323 Harry Hines Blvd., Dallas, TX 75390, USA

Tel: 614-282-0859

Email: Kang-Hsin.Wang@UTSouthwestern.edu

**Funding Statement**

The authors acknowledge the funding support from Cancer Prevention and Research Institute of Texas, RR200042.

**Data Availability Statement**

Data are available from the corresponding author upon reasonable request.

**Acknowledgements**

The authors would like to thank Dr. John Wong from Johns Hopkins University for his input regarding the dual-tube system.

## Abstract

**Background:** A kV dual-tube system has been disseminated as a commercial research platform for preclinical FLASH radiotherapy (RT). Because the tubes are arranged in a parallel-opposed geometry, both output symmetry and time-resolved tube synchronization are critical for achieving sufficiently high dose rates (DR) and reproducible study results. We quantified tube output asymmetry observed in depth-dose measurements as well as tube synchronization and evaluated their impact on FLASH studies.

**Methods:** The dual-tube system defines dose-per-pulse as the combined single-pulse output from both tubes, with DR given by dose-per-pulse/pulse length. 3D dose distributions were reconstructed from film measurements to assess the impact of output discrepancies. Pulse synchronization between tubes was characterized using a scintillator with 1-ms resolution.

**Results:** We showed >20% discrepancies in output at nominally equal mA/ms settings. After we compensated such discrepancy by decreasing the current of the tube with higher output, the inter-tube output difference was reduced to <1%, restoring symmetrical depth dose. We further simulated an in vivo intestinal irradiation in which naïve tube settings resulted in >22% of the organ volume receiving >102% of the prescribed dose, compared with <7% when output compensation was applied. We identified a 10.2 ± 7.0ms synchronization jitter between tubes, which disproportionately impacts the DR at low dose-per-pulse settings, particularly relevant for fractionated studies. Corresponding quality assurance (QA) was designed to monitor tube synchronization over time.

**Conclusion:** We quantified the dosimetric impact of asymmetric output and synchronization and demonstrated implications for preclinical studies. The proposed methodology and QA would mitigate and monitor these effects, ensuring study reproducibility.

## Introduction

Ultra-high dose-rate (≥40 Gy/s, UHDR) radiotherapy (RT) has shown encouraging preclinical results in normal tissue sparing with comparable tumor control relative to conventional dose-rates (~0.1 Gy/s, CONV) irradiation[1-6]. While the majority of RT delivered worldwide is in a photon modality, preclinical FLASH studies have largely focused on electron and proton RT. This emphasis primarily reflects the technical challenges of achieving sufficiently high dose-rates (DR) at research-relevant depths using photon beams[7].

To address this gap, a dual kV X-ray tube system configured in a parallel-opposed geometry has been proposed as a preclinical FLASH research platform, capable of simultaneous single-pulse delivery and achieving a combined DR ≥80 Gy/s at a depth of 10 mm in a water-equivalent medium[8-10]. Demonstrations of the FLASH effect in mouse skin, heart and eye using such devices have been reported[10-13], making it a promising avenue for further investigations of FLASH radiobiology and beam parameters that maximize the FLASH effect. Such work requires precise dosimetry and control of beam parameters including dose, instantaneous and mean DRs, and dose-per-pulse, all of which must be reported to enable accurate comparisons between FLASH and CONV studies and to ensure cross-institutional reproducibility[14].

Using two radiation sources for UHDR irradiation introduces additional dosimetric challenges, since the parallel-opposed geometry demands output symmetry and time-resolved synchronization to achieve high DR and reproducible results. Current literature on dosimetric commissioning of dual X-ray tube devices assumes a high degree of symmetry in output of the two tubes[15,16]. However, from our experience and vendor communication, the output can be tube-specific, in our case with > 20% tube output discrepancy at equal beam current settings. Moreover, the previous studies do not include detailed measurements of dual-tube synchronization, a key consideration in characterizing of effective DR. Here, we present a new dosimetric commissioning procedure that explicitly accounts for tube output asymmetry and pulse synchronization by using film dosimetry and a rapid-response scintillator detector[17], respectively, thereby quantifying the dosimetric profiles and DR delivered by the device. Using an example whole-abdomen intestine treatment plan, we demonstrate that failing to account for output asymmetries of the two tubes can lead to a large fraction of the target volume exceeding the prescribed dose, and correcting the issue leads to a more homogeneous plan.

Crucially for FLASH studies, we measured the effective DR, which is limited by both maximum tube output and the temporal overlap of pulses from each tube. We analyzed beam-on-time mismatch or synchronization jitter between tube pulses and assessed the resulting DR variability. This jitter can decrease the true DR relative to the nominal value and introduce substantial pulse-to-pulse DR variations, with greater impacts for shorter pulses. We therefore provide recommendations for commissioning, quality assurance (QA) and dose delivery protocols of dual X-ray tube FLASH system to mitigate dosimetric uncertainty from output discrepancy and to accurately monitor delivered DR.

## Materials and methods

### X-ray FLASH system

The CIX-FLASH irradiator (Xstrahl Ltd, GA, USA) consists of two independently powered X-ray tubes arranged in a parallel-opposed geometry (**Fig. 1a1-2**) to offset the heel effect and compensate for the rapid dose-rate drop-off with depth[8]. The bottom tube is stationary, whereas the top tube is movable along the Z-direction over a range of 800 mm. Each tube operates at up to 150 kVp, with beam current adjustable from 10 to 630 mA and pulse widths ranging from 10 to 200 ms at 630 mA. Two overhead line laser modules are used for geometric center alignment (X, Y = 0, 0), which is established at the midpoint of the bottom tube window. An additional pulley-mounted laser line module moves with the top tube in a 1:2 ratio and tracks the midpoint between the two tubes (Z = 0). The mouse bed is mounted on a 3D translational micrometer stage for precise placement between the tubes.

### Beam Data, Film Dosimetry and Effects of Output Asymmetry on Dose Distribution

Gafchromic EBT-4 film (Ashland, NJ, USA) was used in solid water (Sun Nuclear SolidWater HE, FL, USA) to measure the absolute dose, depth dose and profile along the depth. Gafchromic films have shown to be dose-rate independent[18] and suitable for the UHDR dosimetry. The film was calibrated at the University of Wisconsin–Madison Radiation Calibration Laboratory (UWRCL, Madison, WI, USA) using the NIST traceable 120M-UW over a dose range of 0–10 Gy. The UW-120M had the closest mean energy to the 150 kVp beam of the FLASH x-ray system (56.6 vs 54.9 keV)[15].

A vendor-provided 3D-printed dosimetry jig[16] was used to reproducibly stack film along the central beam axis (CAX) within solid water at 5 mm increments (60 mm × 60 mm × 5 mm per solid water slab at Z = −10, −5, 0, 5, and 10 mm; **Fig. 1a1**), with a 30 mm tube separation. The source-to-surface distance (SSD) from both tubes to the phantom was approximately 61 mm. The flange serves as the primary collimator and no additional collimators were used (open field). The measurement is up to depth of 10 mm relative to each tube. Films were positioned orthogonally to and centered on the CAX. The X and Y laser lines were marked on the film to precisely locate the geometric center along the stack, which was used as a surrogate for the CAX. In-house software was used to read the laser markers and align the beam profiles measured at different depths. The films were scanned 24 hours post irradiation using a flatbed scanner (Epson Expression 12000XL, Nagano, Japan). An in-house film analysis software utilizing a dual-channel method (green and blue) was used to extract the dose information from films[19,20].

To quantify the impact of output discrepancies beyond traditional depth dose and profiles, we reconstructed the 3D dose distribution by combining depth-dependent film-measured beam profiles using a cubic interpolation algorithm (MATLAB, MathWorks, MA, USA). We further showed a whole-abdomen intestinal treatment plan using a published mouse CT dataset[21] as an example to illustrate the impact on intestinal dose volume histograms (DVHs) with and without tube output compensation.

### Temporal Beam Characterization with Scintillator

Temporal pulse characteristics and jitter between pulses were assessed using a scintillation detector (HYPERSCINT RP-FLASH, Medscint, Quebec, Canada) with 1-ms temporal resolution[17], well under the minimal available 10 ms pulse length. The scintillator was placed in the beam path closer to the bottom X-ray tube to distinguish the signal from the two tubes by preferentially amplifying the bottom-tube signal. The beam-on threshold was defined as 50% of the peak amplitude for each pulse. The scintillator measurements were used exclusively for relative-intensity pulse width measurement so no dosimetric characterization was needed.

## Results

### Correcting Depth Dose Asymmetry via Tube Output Compensation

We first analyze the depth dose of the two tubes along the CAX. With identical 100 ms and 630 mA setting for both tubes, the depth dose curves show asymmetrical distribution (**Fig. 1b**). If the depth dose of the bottom tube is flipped, it reveals a consistent ratio of $1.28 \pm 0.02$ between the outputs of the top and bottom tubes throughout the depth (**Fig. 1c**). This output discrepancy may be compensated by decreasing the beam current of the tube with higher output. In our case, by setting 500 mA to the top tube and 630 mA to the bottom tube (100 ms for both), the output difference between the two tubes is reduced to within 1% and symmetrical depth dose are restored (**Fig. 1d, e**). It is worth noting that this approach can reduce DR, by ~10% in our case, as indirectly indicated by the decrease in depth-dose (**Fig. 1 b vs. d** ).

**Figure 2** further illustrates the dose and DR profiles in an open field for each tube (top **Fig. 2a-c,** bottom **Fig. 2d-f**) and the combined dual-tube configuration (**Fig. 2 g-i**) at various depths, all acquired using nominally identical output settings (100 ms, 630 mA). Each tube exhibits the heel effect along the inline X direction (**Fig. 2a-f**). Consistent with **Fig. 1b**, the higher output of the top tube produces a stronger heel effect at shallow depths (**Fig. 2a vs. d**) and contributes more to the 2D DR distributions at the midpoint between the two tubes Z = 0 (**Fig. 2b vs. e**). While the parallel-opposed geometry improved the symmetry of the composite profile at the midpoint (**Fig. 2g-h**), the profile remains asymmetric away from the midline, where it is dominated by the proximal tube (e.g., Z = - 10 mm, **Fig. 2i**).

**Effects of Output Asymmetry on 3D Dose Distribution and Evaluation of Output Compensation Correction**

We further reconstructed the 3D dose distribution within a 40 × 40 × 20 $mm^3$ by interpolating 2D depth-resolved dose profiles. With asymmetric tube output, the dose distribution within a 20 x 20 x 14 $mm^3$ region of interest (ROI), representative of a typical mouse irradiation geometry, was likewise asymmetric, with notable hotspots towards the higher output (top) beam (**Fig. 3a1, 2**). In contrast, using the output-compensated method (**Fig. 1d,e)**, a more homogenous irradiation was achieved within the ROI, with the majority of volume within 5% of 100% isodose line (**Fig. 3b1,2 vs a1,2**).

Next, we illustrated the effects of our corrective strategies using an actual mouse intestine irradiation plan as an example (**Fig. 3 c1,2**). In a typical setup, the combined thickness of irradiated mouse tissue and the ~2.5 mm thick water-equivalent mouse bed is ~20 mm, allowing the dose distribution measured in a 20-mm-thick phantom (Fig. 3a, b) to serve as a reasonable approximation for the intestinal case. The 20 x 20 x 14 $mm^3$ ROI matches well with the dimensions of a mouse intestine (**Fig. 3c1,2,** red ROI). In the naïve configuration, the intestinal DVH demonstrates substantial dose inhomogeneity, with >22% of the organ volume receiving >102% of the prescribed dose, and 7% of the volume receiving >105% of the dose (**Fig. 3d**, black dash line), primarily due to output and profile asymmetry between the tubes. Applying the output compensation markedly improves dose uniformity, reducing the volume receiving >102% dose to below 7%, and that receiving >105% dose to below 0.1% (**Fig. 3d**, blue line). However, in our system, the maximum DR decreases from 82 Gy/s (uncompensated) to 72 Gy/s after compensation.

**Dose Rate Uncertainty and Reduction from Tube Synchronization Mismatch**

We characterized the beam-on time accuracy at 1 ms temporal resolution for both single-tube and dual-tube configurations using the scintillator (**Fig. 4a**). For either the top or bottom tube operated individually, the measured pulse length agreed with the nominal setting within the 1 ms temporal resolution limit (**Fig. 4b**). However, a beam-on time mismatch (or jitter) was observed in the dual-tube configuration (**Fig. 4c**), defined as $\Delta T = T_{start}^{top} - T_{start}^{bottom}$, which refers to the time difference observed between the top and bottom tube beam onset. The jitter followed a normal distribution and was independent of the nominal pulse length setting, as $\Delta T$ was not significantly different between the 100 and 200 ms settings ($p = 0.58$) (**Fig. 4 d vs e**). The absolute value of average jitter $|\Delta T|$ is 10.2 ± 7.0 ms (**Fig. 4e**) when combining the data from both 100 and 200 ms pulse length settings.

The beam-on time jitter leads both to DR reduction and uncertainty. Because the effective pulse duration is prolonged by $|\Delta T|$, the actual average DR decreases. In addition, the random nature of $|\Delta T|$ introduces variability in the delivered DR. The actual average DR is defined as $\dot{D}_a = \frac{D}{T+|\Delta T|}$ whereas the nominal DR defined as $\dot{D}_n = \frac{D}{T}$, where $D$ is the dose and $T$ is the nominal pulse width. We assumed users prefer symmetrical output, so we quantified DR uncertainty and reduction based on the output-compensated case. In **Fig. 5a**, the dashed line represents the nominal DR, the solid line represents the actual average DR, and the shaded region indicates the associated DR uncertainty. Obviously, both the DR reduction and uncertainty are more pronounced at lower single-pulse doses. Furthermore, these effects become more substantial at higher beam currents, corresponding to high nominal DR irradiation conditions. We further present color plots that provide practical guidance on the percentage reduction in actual DR and the associated DR uncertainty for a given dose and DR combination, shown in **Fig. 5b** and **c**, respectively. For example, given the measured jitter in our system, a single 14.4 Gy pulse delivered at a nominal dose rate of 72 Gy/s can be achieved with less than 5% DR reduction and uncertainty. In contrast, delivering the same total dose as 4 × 3.6 Gy pulses or 3.6 Gy/fraction at the same nominal DR results in an average delivered DR of only ~82% of the nominal value (**Fig. 5b**), with > 15% DR uncertainty across individual pulses (**Fig. 5c**). This indicates that for UHDR delivery, lower prescribed doses result in larger DR reductions and greater DR uncertainty (**Fig. 5a** red and blue line, **b** and **c**).

## Discussion

This work has particular significance because the dual-tube kV X-ray system evaluated here is a commercially available platform that is increasingly adopted across institutions for preclinical FLASH research. Unlike single-site experimental systems, commercial dissemination implies that dosimetric characteristics such as tube-specific output asymmetry and beam-on-time synchronization jitter will directly affect the reproducibility and comparability of FLASH studies across laboratories.

Our results show that tube output discrepancies significantly affect the depth dose curve and can lead to asymmetrical distributions (**Fig. 1b,c**). To dosimetrically address the issue of beam output difference, we recommend users perform individual and combined tube depth-dose measurements (**Fig. 1b,c**). If asymmetrical depth-dose distributions are observed, the asymmetry may be resolvable by matching the beam outputs (**Fig. 1d,e**). Because such output variation can be tube-specific[15,16], users should discuss machine-specific solutions with the vendor. The main drawbacks of the output-compensation approach are the reduced maximal DR resulting from tuning one of the tube outputs down and the possibility that the limited discrete tube-current presets may not allow exact tube output matching in every case. This finding suggests that the tube current presets defined by the manufacturer (i.e., 630, 500, 400, etc. mA) should be revised to allow for higher-resolution adjustment of the tube current.

The dose heterogeneity arising from the asymmetrical output would particularly affect irradiations with large ROI, such as abdomen. We show that the proposed corrective approach enables whole-abdomen irradiation with relative homogenous intestinal dose, achieving 95-105% isodose coverage (**Fig. 3b vs. a**). To ensure long-term stability, a routine QA should be established to verify the output and beam quality of each tube. More comprehensive evaluations, such as full 3D dose profiling, may be reserved for annual QA. Owing to the lack of an appropriate treatment planning system, we acknowledge the dosimetric limitations of using solid-water measurements (**Fig. 3**) to infer in vivo dose distributions. Integration of a treatment planning system incorporating accurate X-ray source models and system geometry would further enable precise UHDR dosimetry and support in vivo FLASH investigations.

We also demonstrate that pulse-timing jitter between the two tubes leads to both variability and undershoot of the DR, with disproportionately larger effects at low dose-per-pulse under high nominal DR settings (**Fig. 5a**). It has been suggested that there is a minimal threshold dose below which the FLASH effect may be absent[22]. This further implies that FLASH-RT may not be appropriate for standard fractionation, such as 2 Gy/day treatment schedules. Considering the shift in our field towards short-course, high dose regimens, investigating FLASH-RT within hypofractionated treatments is essential for successful clinical translation[23]. Recent works indicate that fractionation can markedly reduce the FLASH effect[12,24,25], with substantial reductions even when splitting the dose across multiple pulses in a single session[26]. Precise DR control and accurate reporting are therefore crucial for ensuring the reproducibility of FLASH studies[27], particularly given conflicting results reported in the field and the urgent need to determine its clinical translatability. In our system, a nominally 72 Gy/s irradiation delivered in a single 200 ms pulse would result in an actual DR of 68 ± 3 Gy/s, but the same cumulative dose delivered using the same nominal DRs (i.e., same tube current settings) split across 4x50 ms pulses would result in an actual DR of 60 ± 10 Gy/s (**Fig. 5b,c**) with as high as 17% DR variation. This implies that, for hypofractionated studies using the dual-tube X-ray FLASH system, larger DR uncertainties are expected. Delivering consistent actual DRs across pulses of different durations therefore requires careful selection of nominal DRs that compensate for tube-timing jitter. Accordingly, measuring $|\Delta T|$ during commissioning and monitoring it as part of routine QA are strongly recommended.

To characterize the jitter, we recommend performing repeated scintillator measurements with both tubes as part of commissioning and annual QA, with the measurement setup and analysis similar to those shown in **Fig. 4**. The number of samples $N_0$ needed to generate a representative distribution of the timing jitter in the commissioning measurement depends on the desired precision. In a previous study, Mansoury et al. provide an order-of-magnitude estimate of the jitter to be ~ 10 ms, in agreement with our measurements[16]. This value can be used as an initial approximation for the mean and standard deviation of $|\Delta T|$, from which the expected uncertainty of the mean value of $|\Delta T|$ from $N_0$ measurements can be approximated as $\pm\frac{20\,ms}{\sqrt{N_0}}$. With $N_0$ = 100 measurements, this uncertainty becomes ± 2 ms, significantly smaller than a typical pulse duration, on the order of ~100 ms, and comparable to the resolution of the RP-FLASH scintillator[17]. The mean $\mu_{|\Delta T|}$ and standard deviation $\sigma_{|\Delta T|}$ of the parameter of interest $|\Delta T|$ from such a sufficiently large initial sampling can be then treated as reference values (**Fig. 4f**). From $\mu_{|\Delta T|}$ and $\sigma_{|\Delta T|}$, the accurate expected DRs and their uncertainties are easily determined using $\dot{D}_a = \frac{D}{T+\mu_{|\Delta T|}}$. Our recommendation for monthly QA is to verify consistency of the $|\Delta T|$ parameter. Therefore, we suggest performing the same timing offset measurements as for the annual QA, but with a smaller sample of repeats $N$, which can be adjusted to a desired confidence level. A natural acceptance criterion is that the sample mean $\mu_N$ falls within $\mu_{|\Delta T|} \pm \frac{2\sigma_{|\Delta T|}}{\sqrt{N}}$, as this range covers 95% of expected sample mean distributions. In our system, using $N$ = 5 gives a $\mu_N$ acceptance range of $10 \pm 9$ ms. While such a wide tolerance is appropriate for tracking long term stability of the tube jitter, during treatment we recommend monitoring $|\Delta T|$ using an external detector, such as a scintillator detector placed outside of the treatment field, to further eliminate uncertainty by quantifying the delivered DR.

## Conclusion

Dual-tube kV X-ray irradiators provide an accessible UHDR photon source for preclinical FLASH studies. We demonstrate that simple adjustment of the beam current can compensate for the output mismatch between the two tubes, thereby improving dose homogeneity. Tube beam-on time jitter leads to dose-rate reduction and increased dosimetric uncertainty, which can be quantified using a scintillation detector. Our work provides practical methods for characterizing, mitigating and monitoring beam delivery uncertainties in the dual-tube X-ray FLASH system, ensuring study reproducibility.

## Figures

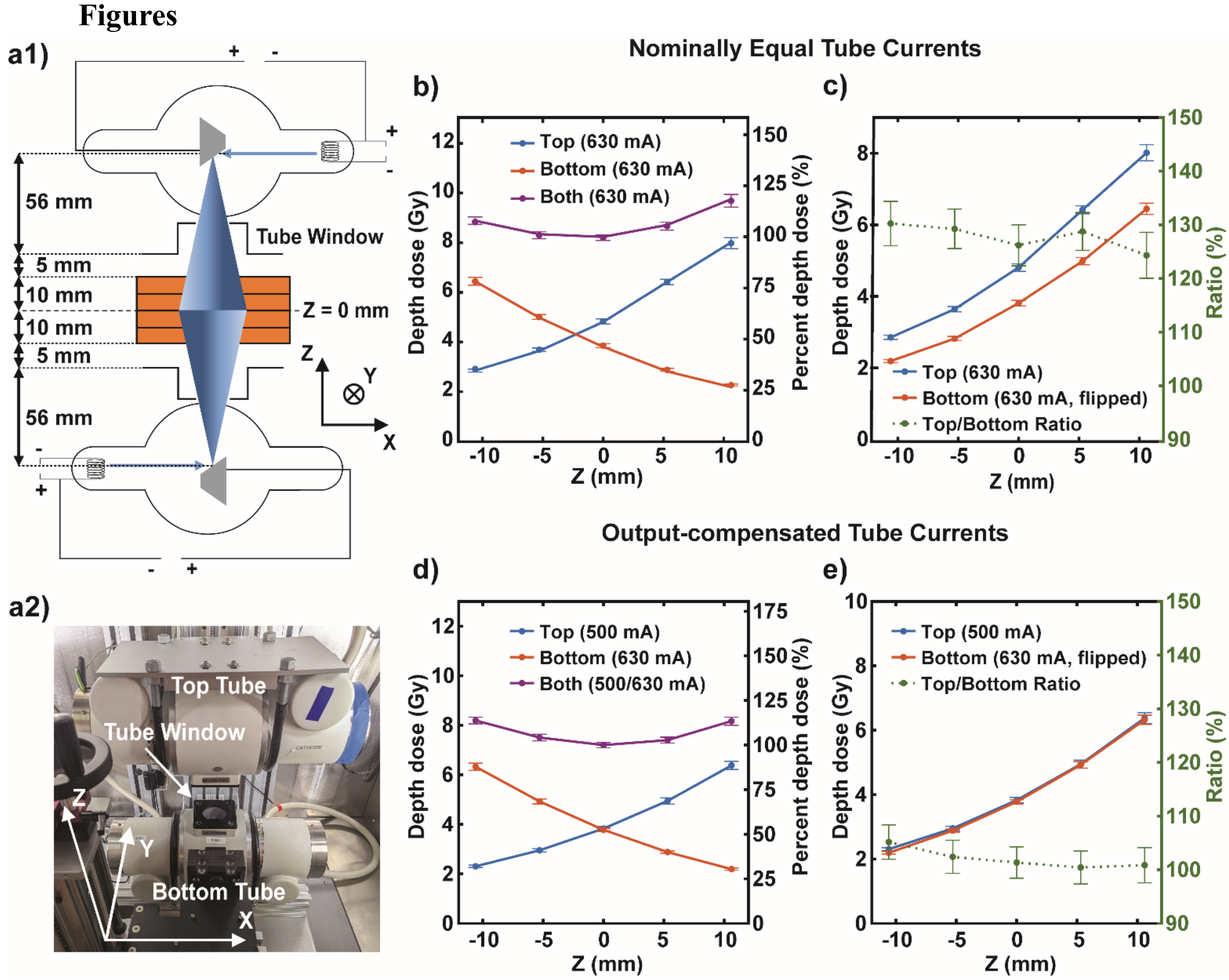

**Figure 1. Single- and dual-tube depth-dose measurements.**
**a1)** Diagram of a commissioning dosimetry setup for the dual tube X-ray system (not to scale) with a solid water stack (orange). Blue color gradient indicates lateral variation in beam intensity from each tube due to the heel effect in the inline (X-axis) direction. **a2)** Image of the actual system. **b**) Depth dose for nominally equal output tube currents. Right axis normalized to total dose delivered using both tubes at Z = 0 mm. **c)** Comparison of dose curves for individual tubes from b), with the bottom tube curve Z-axis inverted to facilitate direct comparison of depth curves. Right axis and green data show the top/bottom ratio of the two depth curves. **d)** and **e)** are the figures corresponding to b) and c), respectively, for the case of output-compensated tube currents.

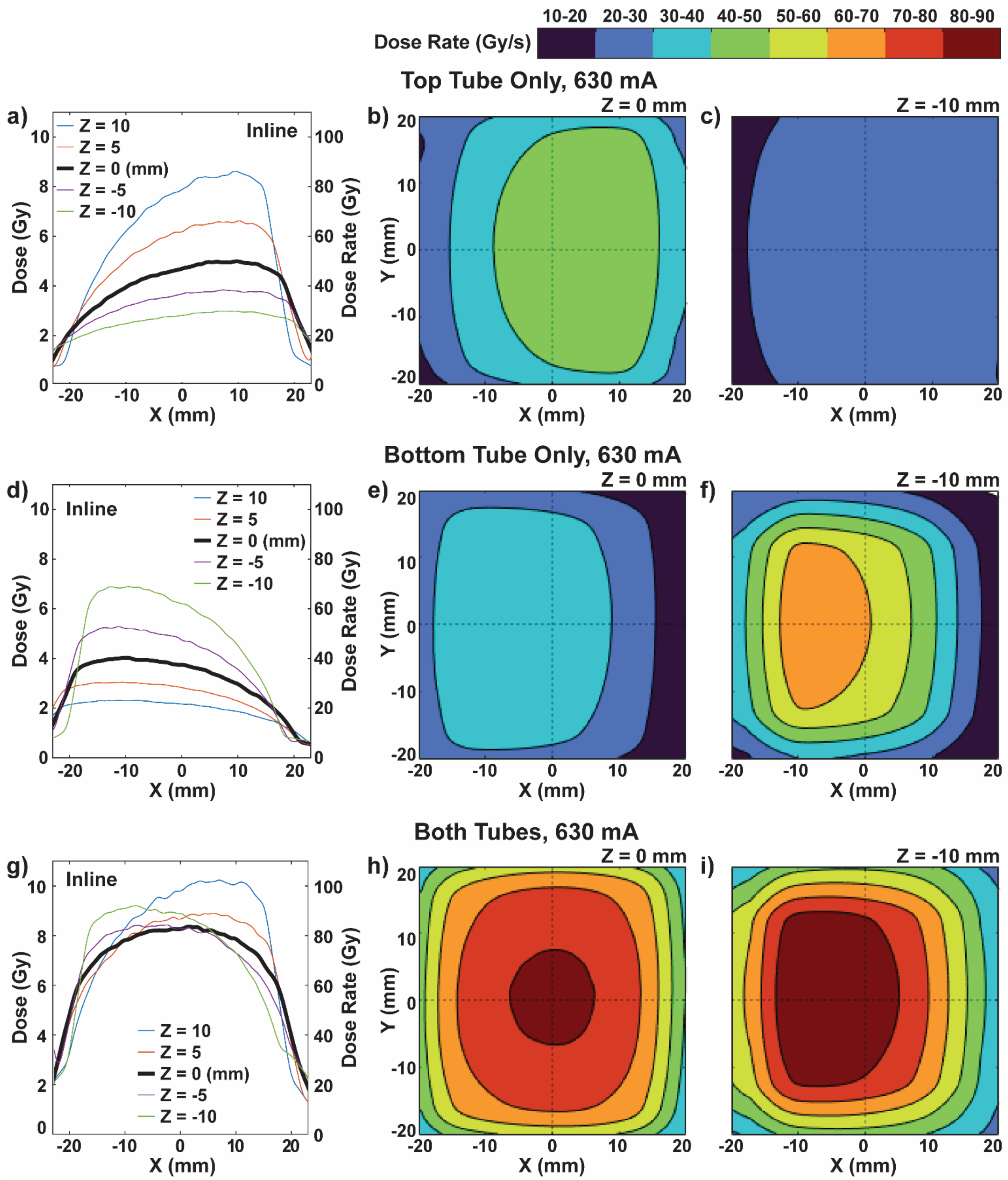

**Figure 2. Single- and dual-tube beam profiles.**
**a)** Inline (X-axis) beam profiles through the mechanical center axis (Y = 0 mm), using only the top tube. **b)** 2D beam profiles at vertical beam center (Z = 0 mm) using only the top tube. **c)** 2D beam profiles close to the bottom tube (Z = -10 mm) using only the top tube. **d-f)** and **g-i)** are the figures corresponding to a, b, c) for top and both tubes, respectively. Both tubes were set to 630 mA, 100 ms for all subpanels.

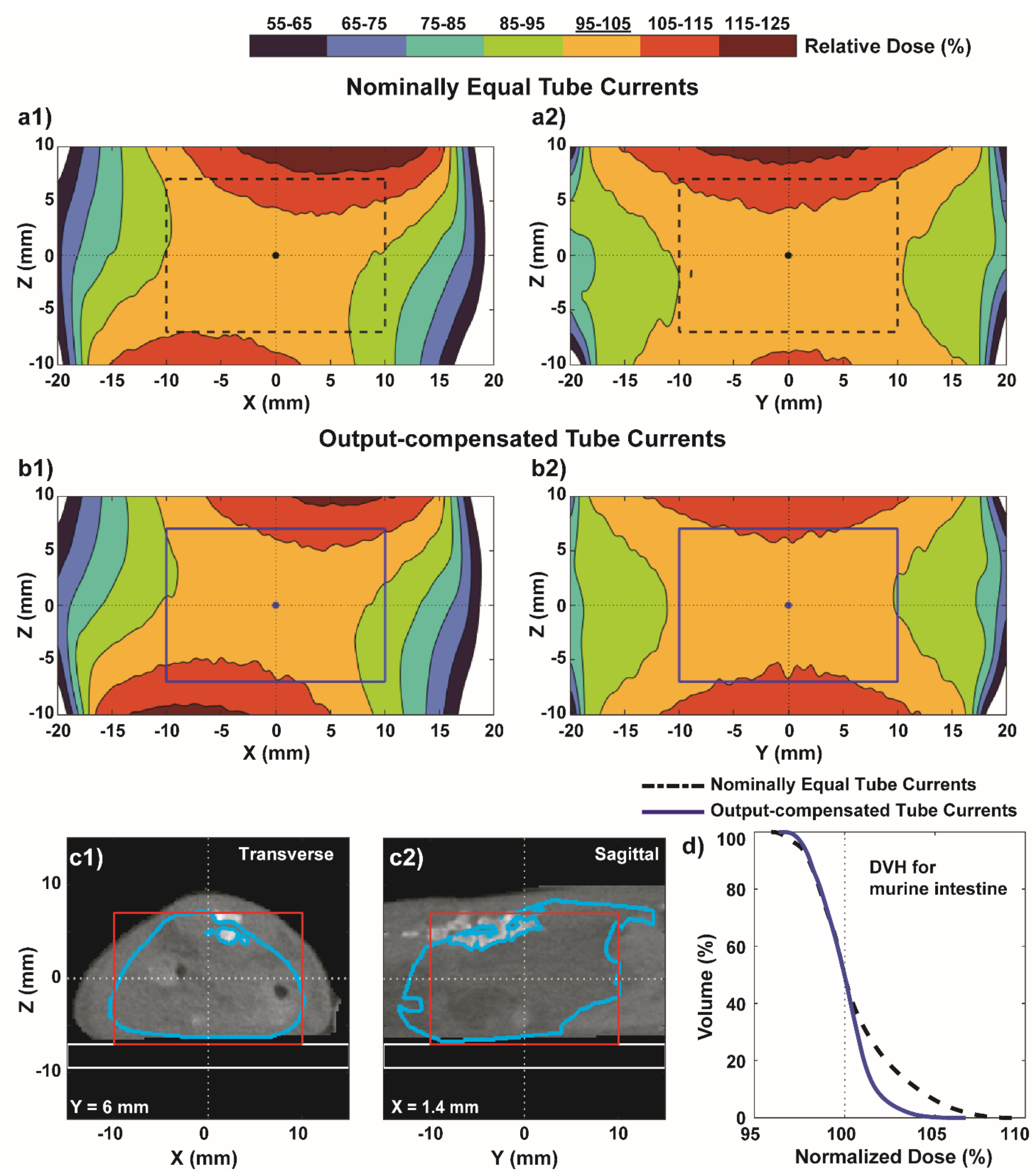

**Figure 3. Correcting non-uniform dose delivery with output compensation.**
**a)** Inline **a1**) and cross-line **a2**) profiles along the beam axis, using a nominally equal 630 mA, 100 ms setting for both tubes. Full 3D profiles were reconstructed from 2D film-measured profiles measured at Z = -10,-5,0,5,10 mm positions. A 20 x 20 x 14 mm$^3$ ROI centered on the mechanical center (X, Y, Z = 0 mm) is shown (black dash-dot line). Colorbar shows relative dose normalized to the dose measured at the mechanical center. **b)** Same as (a), but using output-compensated settings (630 mA, 100 ms for the bottom and 500 mA, 100 ms for the top). A 20 x 20 x 14 mm$^3$ ROI centered on the Z = 0 mm is indicated by the blue solid line. **c)** CT image of a mouse abdomen with a 20 x 20 x 14 mm$^3$ ROI in a transverse **c1**) and sagittal **c2**) view. The segmented intestine region is outlined in light blue. **d)** DVHs for the intestine using different beam configurations from a) and b); the DVHs were normalized to dose at 50% volume for the sake of comparison.

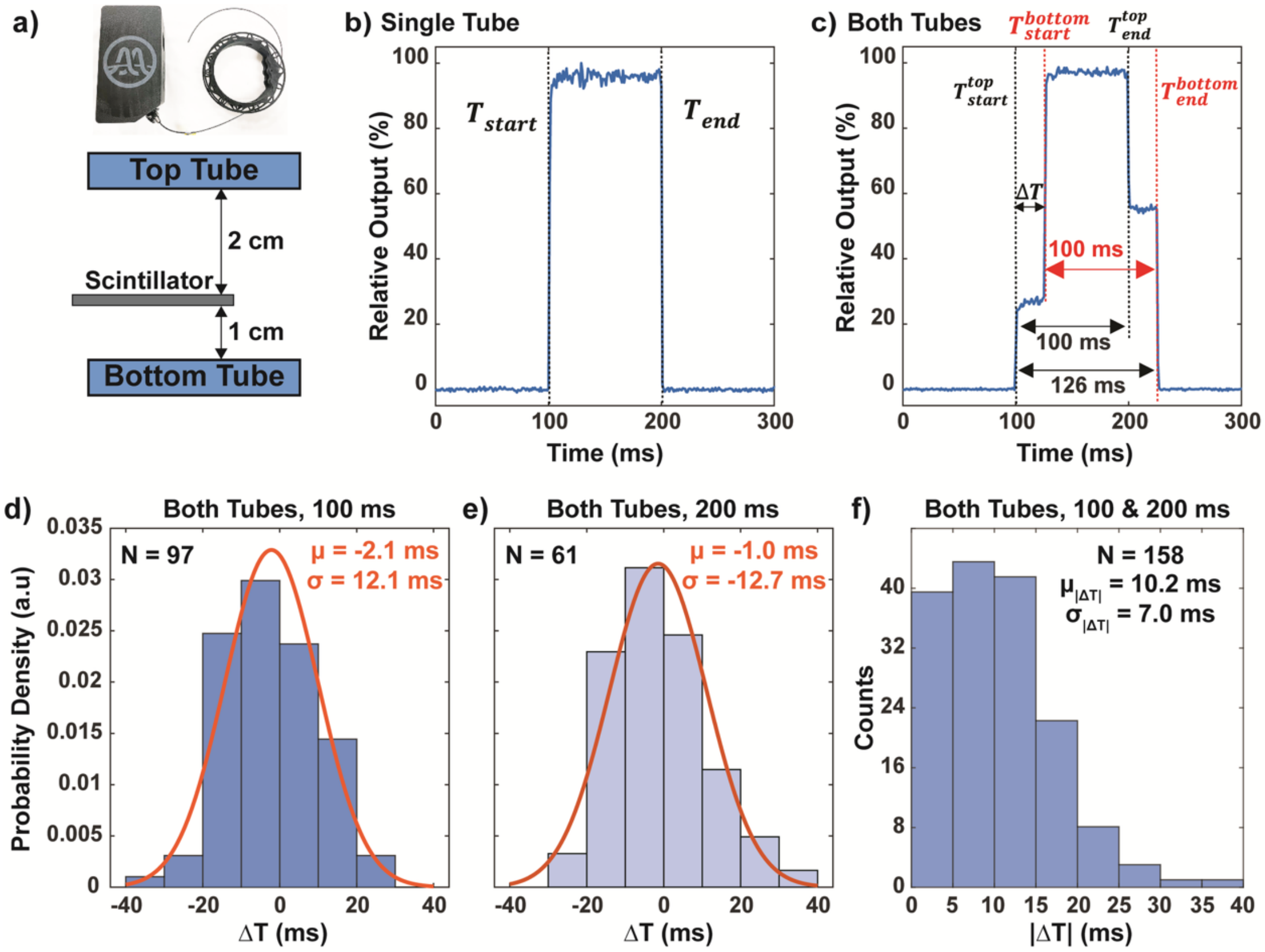

**Figure 4. Scintillator measurements of tube pulses.**
**a)** Diagram of beam timing measurements using a scintillator. The scintillator was placed closer to the bottom tube, to distinguish the bottom and top tube pulses. **b)** A 100 ms pulse from a single tube. $T_{start}$ and $T_{end}$ are defined as the time when pulse reaches 50% of maximum output. **c)** One instance of a pulse produced by both tubes. While both pulses are individually 100 ms, the offset between the start time of the top and bottom pulses $\Delta T$ increases the total pulse length. **d)** Histogram of $\Delta T$ using 100 ms pulses with Gaussian fit (red). **e)** Histogram of $\Delta T$ using 200 ms pulses with Gaussian fit (red). **f)** Combined histogram of $|\Delta T|$ using 100 and 200 ms irradiation.

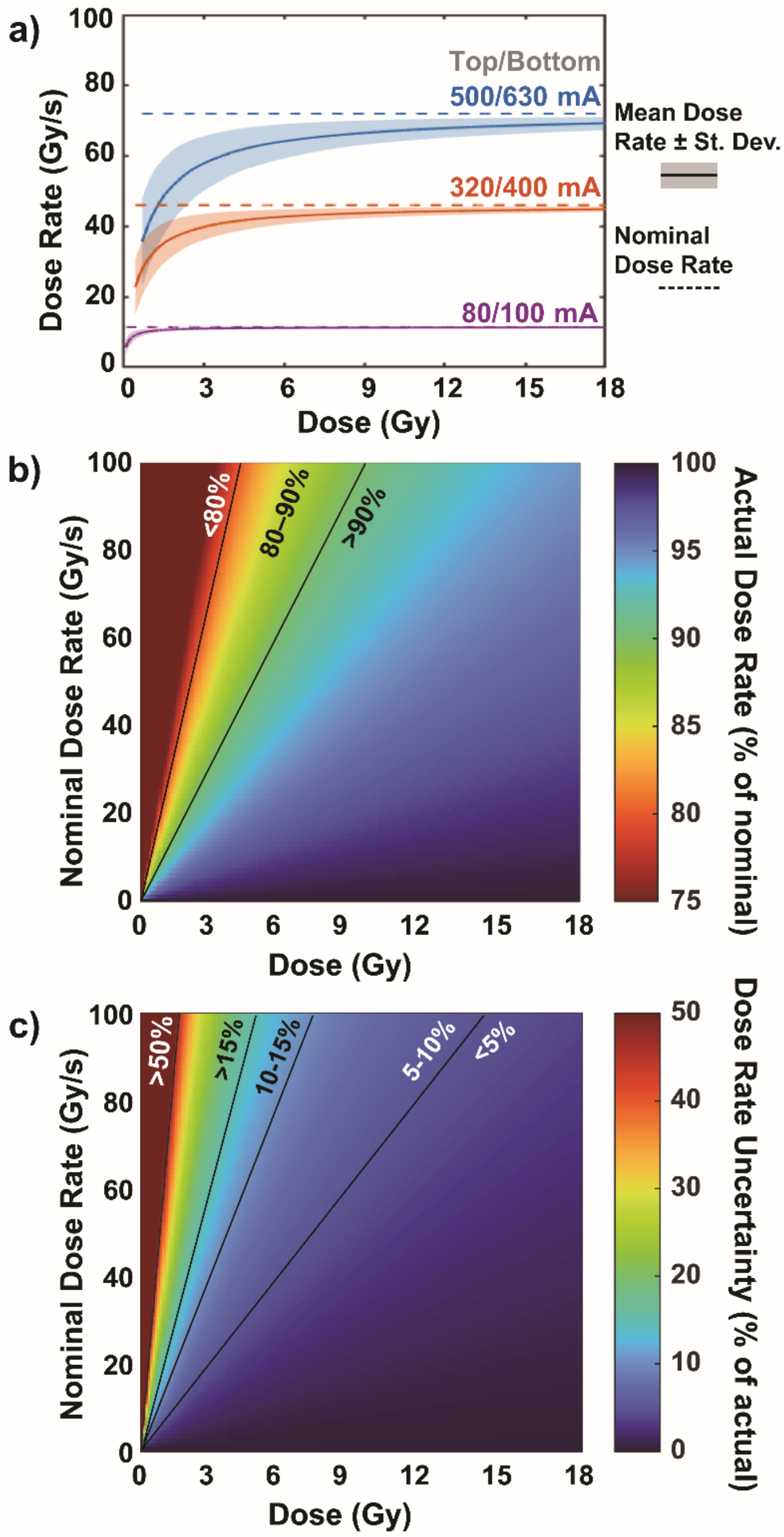

**Figure 5. Tube jitter decreases dose-rate and increases dose-rate uncertainty.**
**a)** Comparison of nominal and actual dose-rates vs doses for various output settings. Smaller doses delivered at higher dose-rates suffer greater dose-rate undershoot and uncertainty. **b)** Color plot of actual dose-rates (expressed as % of nominal) as a function of nominal dose-rate and dose. **c)** Color plot of relative dose-rate uncertainty as a function of nominal dose-rate and dose.